\documentclass[twocolumn,aps,preprintnumbers,amsmath,amssymb]{revtex4}
\usepackage{graphics}
\usepackage{graphicx}
\usepackage{dcolumn}
\usepackage{bm}
\usepackage{epsfig} 
\usepackage{psfrag} 

\usepackage[dvips]{color}

\newcommand{\beq}{\begin{equation}}
\newcommand{\eeq}{\end{equation}}
\newcommand{\D}{| \psi |}
\newcommand{\Db}{| \bar{\psi} |}
\newcommand{\Do}{| \psi_0|}

\newcommand{\R}{| \tilde \psi |}

\newcommand{\lt}{\left}
\newcommand{\rt}{\right}

\newcommand{\smallequa}{\ \lower-1.2pt\vbox{\hbox{\rlap{$<$}\lower5pt\vbox{\hbox{$\sim$}}}}\ }

\begin{document}




\title{Thermodynamics and Phase Diagram of High Temperature Superconductors.}

\author{Philippe Curty and Hans Beck}
\affiliation{ Universit\'e de Neuch\^atel, 2000 Neuch\^atel, Switzerland}
\begin{abstract}
 Thermodynamic quantities are derived for superconducting and pseudogap regimes by taking into account both amplitude and phase fluctuations of the pairing field.
    In the normal (pseudogap) state of the underdoped cuprates, two domains have to be distinguished: near the superconducting region, phase correlations are important up to the temperature $T_\phi$. Above  $T_\phi$, the pseudogap region is only determined by amplitudes, and phases are uncorrelated.
   Our calculations show excellent quantitative agreement with specific heat and magnetic susceptibility experiments on cuprates. We find that the mean field temperature $T_0$ has a similar doping dependence as the  pseudogap temperature $T^*$, whereas the pseudogap energy scale is given by the average amplitude above $T_c$.
\end{abstract}
\maketitle
One of the most intriguing problems in high temperature superconductivity is the presence of a region above the critical temperature $T_c$ and below a temperature $T^*$ where observable quantities deviate from Fermi liquid behaviour. This region is called pseudogap region \cite{loram94,timusk} because it contains effects similar to superconductivity like a partial suppression of electronic density of states.


The origin of such a pseudogap above $T_c$ is unclear. There are four major approaches
concerning its theoretical understanding: the first is based on the formation of
incoherent Cooper pairs above $T_c$. Phase order \cite{emery} or Bose
condensation \cite{randeria} would then establish superconductivity at $T_c$. The second
assumes that the pseudogap is induced by anti-ferromagnetic fluctuations \cite{schmalian}. The third approach is based on spin-charge separation where spins bind together to form
spin-singlets and the energy needed to split them apart leads to the formation of a "spin-gap"
\cite{lee}.
 The fourth assumes the existence of a quantum critical point \cite{sachdev} but the latter has never been observed. However, these approaches seem to be unable to describe specific heat and magnetic susceptibility.

\begin{figure}[h]
\begin{center}
 \setlength\unitlength{1cm}
\begin{picture}(8,8)
\put(0,0){\resizebox{8cm}{!}{\includegraphics{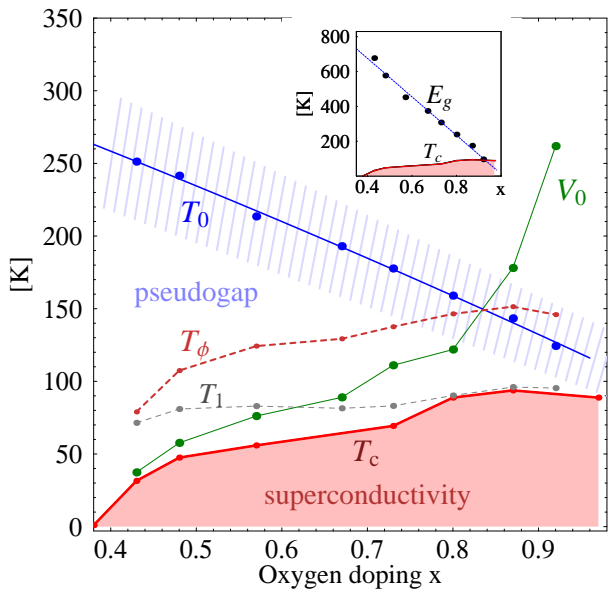}}}
\end{picture}
\end{center}
\vspace{-0.5cm}
\caption{ Phase diagram of YBa$_2$Cu$_3$O$_{6+x}$. Effects of amplitudes are large in the quasi uncorrelated
 pseudogap region (below $T_0$), whereas phase correlations remain important only below $T_{\phi}$. 
  The temperature $T^*$, where $\gamma$ and $\chi_d$ cross over to normal behaviour, is located in
   the hatched area. {\em Inset:} the pseudogap energy scale $E_g$.}
\label{phasediagram2}
\end{figure}

The main aim of this article is to show that various experimental observations
can indeed be interpreted in terms of fluctuations of the pairing field $\psi = |\psi| e^{i \phi}$, and that
two temperature regions have to be distinguished (see Fig. \ref{phasediagram2}): for a
relatively small temperature interval $T_c < T < T_{\phi}$ the phase of  $\psi$ is still
correlated in space over
some correlation length $\xi$ (the Kosterlitz-Thouless correlation length in $2D$).
 Thus, in this regime, observables are
governed by correlated phase fluctuations described by the XY-model. For $T_{\phi} < T <
T^*$, phases of $\psi$ are essentially uncorrelated ($\xi$ is on the order of the
lattice constant), but $\D$ is still fluctuating and non-zero, signaling local pair fluctuations. 
This explains the wide hump seen in specific heat experiments \cite{loram94}, the
depression of the spin susceptibility \cite{takigawa} and the persistence of the pseudogap for
$T < T^*$.

Our method has a major difference with respect to the Emery and Kivelson phase fluctuations scenario \cite{emery} of the pseudogap regime: we show that phase fluctuations influence the pseudogap only up to a temperature  $T_{\phi} $ which is much smaller than $T^*$. Above $T_{\phi}$, observables are thus only determined by the amplitude of the pairing field.


The picture of two different regimes above $T_c$ is also supported by other experiments: {\bf a)} Demsar {\em et al} \cite{demsar}, interpreting the real-time measurements of the
quasiparticle relaxation dynamics, find a temperature interval of only a
 few K in which pair fluctuations associated with their collective phase are important,
whereas the pseudogap persists to much higher temperatures.
{\bf b)} Hall effect measurements \cite{matthey} in underdoped GdBa$_2$Cu$_3$O$_{7-x}$
show a characteristic temperature $T'$ between $T_c$ and $T^*$, at which the temperature
dependence of $\cot(\Theta_H)$ deviates from $T^2$ and the Hall coefficient has a peak. A possible explanation consists in considering vortex excitations as scattering centers
modifying the Hall angle $\Theta_H$ and Hall coefficient, which would again suggest
identifying $T'$ with our $T_{\phi}$ below which correlated vlatex ortices exist.

We base our calculations on a $d$-wave attractive Hubbard model
\begin{equation}
\label{Hamiltonian}
H = -t \sum_{\lt<i,j \rt> \sigma}\ c^{\dagger}_{i \sigma} c_{j \sigma}
- U \sum_{i} Q_{d}^{\dagger}(i) \ Q_{d}(i)
\end{equation}
with a hopping $t$ between nearest neighbour sites $i$ and $j$ on a square lattice.
The interaction favours the formation of onsite $d$-wave pairs since
$
Q_{d}^{\dagger}(i) =  \sum_j D_{ij} \ Q^{\dagger}_{ij}
$
where $D_{ij}=1, (-1)$ for $i$ being the nearest neighbour site of $j$ in horizontal
(vertical) direction.
$
Q_{ij}^{\dagger} = ( c_{i \uparrow}^{\dagger} c_{j
 \downarrow}^{\dagger} - c_{i \downarrow}^{\dagger} c_{j \uparrow}^{\dagger} )/\sqrt{2}
$
 is the singlet pair operator of neighbouring sites.
Decoupling the interaction with the help of a Stratonovich-Hubbard
transformation, the partition function $Z = \mbox{Tr} \ e^{- \beta H}$ is then
\begin{equation}
\nonumber
 Z =  Z_n \int D^2 \! \psi \lt< {\cal T} e^{-  \int_0^{\beta} d\tau
\sum_{i} \lt(\frac{1}{U} |\psi|^2 +  \psi \ Q_d^{\dagger}(i) + \mbox{\small hc} \rt)} \rt>_{H_n}
\end{equation}
where $\psi = \psi(i, \tau)= |\psi(i,\tau)| e^{i \phi(i, \tau)}$, and $H_n$ is the
non-interacting part. The trace over the fermionic operators can be evaluated yielding
\begin{equation}
\label{Z.fermionic}
Z = \int D^2 \! \psi \ e^{-\int_{0}^{\beta}  d \tau \lt[ \sum_i \frac{1}{U}
|\psi|^2  + \mbox{Tr} \ln G \rt]}.
\end{equation}
Here $G$ is a Nambu matrix of one-electron Green functions for fermions interacting with
a given, space and time
dependent pairing field $\psi(i, \tau) $.

Expanding (\ref{Z.fermionic}) in powers of $\vec{\nabla} \psi$, $Z$ can be written as a
functional integral involving an action $S[\psi]$ for a field $\psi$ that changes
slowly in
space and that can be taken time-independent:
\beq
S[\psi] =  \int d^d \! r \ \left[  S_0(\D) \ + \ S_1(\vec{\nabla}\psi) \right]
\label{action}
\eeq
where $S_0$ is a local function of $|\psi({\bf r})|$, and
$S_1 = c {|\vec{\nabla} \psi |^2 / 2}$ where $c$ is a constant. $d$ is the dimension.\\


Now we would like to compute observables such as energy, specific heat and spin susceptibility. Our  main strategy is to treat separately amplitude and phase fluctuations. In this spirit, two different approaches are possible: 
1) The amplitude is fixed and determined by a suitable variational equation. 
2) The energy is expanded around the average amplitude.


\noindent
{\bf Variational Method:} we can neglect amplitude correlations since simulations show that they are weak between different sites $i, j$: 
$\langle |\psi_i| |\psi_j| \rangle - \langle \D^2 \rangle \approx 0$ 
(i.e. the amplitude is always positive and shows no critical behaviour). Rewriting the free energy $F$ in terms of a constant amplitude $\Db$ yields
\beq
\nonumber
F = -{1 \over \beta} \log  \int D \phi  \ e^{-\beta \lt[V S_0(\Db) - \log(\Db)
V/\beta + \int{d^d \! r \ S_1} \rt] }
\eeq
where the Jacobian of the polar transformation is put into the exponential, and $V$ is the
volume. Equating  the derivative of $F$ with respect to $\Db$ to zero leads to
the self-consistent equation:
\beq
{\partial S_0(\Db) \over \partial \Db } - {1 \over \beta \Db} + c  \Db \ \lt< | \vec{\nabla}
e^{i \phi} |^2\rt>_{XY} = 0.
\label{amplitudeequation}
\eeq
The first term, called the amplitude contribution, leads to the BCS gap equation \cite{BCS} if other
contributions are neglected. The second comes from the Jacobian and  implies that the 
amplitude is never zero. The third term is the expectation value of the energy
in the XY  model depending on the constant coupling $ c \Db^2/2$. This
contribution characterizes the influence of the phase fluctuations. Solutions of equation
(\ref{amplitudeequation}) are reliable for all temperatures except for $T<<T_c$. They are
 accurate at $T_c$ only in the underdoped regime.


\noindent
{\bf Average Value Method:}
now, we keep the {\em local} coupling in $S_1$ between amplitude and phase by introducing the statistical Ginzburg-Landau (GL) model (see \cite{curty}).  Averages like  $\langle \D \rangle$ and $\langle  S_1 \rangle$ are computed within the GL model with all fluctuations and correlations. Then fermionic observables are expanded with respect to the average amplitude.
\begin{figure}[h]
  \begin{center}     
     \setlength\unitlength{1cm}
     \begin{picture}(8,5)
        \resizebox{8cm}{!}{\includegraphics{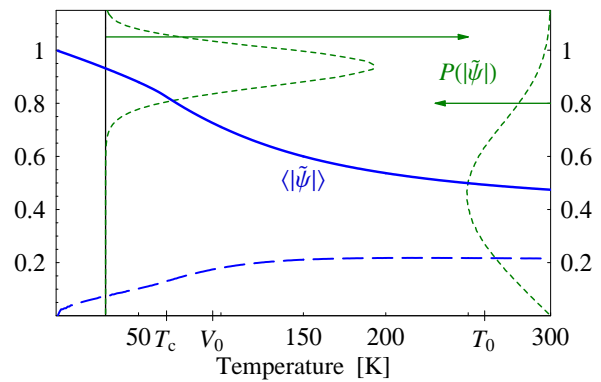}}
     \end{picture}
  \end{center}
\vspace{-0.5cm}
\caption{ Average amplitude $\langle \R \rangle$(thick) and the corresponding standard deviation $(\langle \R^2 \rangle - \langle\R\rangle^2)^{1/2}$ (long dashed) from 2D GL simulations. Parameters are $V_0 = 95$K, $T_0 =260$K. Dashed curves: amplitude distribution $P(|\psi|)$ (histogram) for $T=300$K (right) and $T=30$K (left).}
\label{fig-distribution}
\end{figure}
Expanding the energy density obtained from action (\ref{action}) around the average amplitude (see Fig. \ref{fig-distribution}) yields:
\beq
E = \langle  S \rangle_{S}/V \approx S_0(\langle \D \rangle) + \langle  S_1 \rangle_{S} + {\cal O}(\langle \delta|\psi|\rangle^2) 
\eeq
where the {\em square} of amplitude fluctuations is neglected in first approximation. 
Averages are computed using a normalised GL action:
\beq
 	S_{GL}[\psi] = k_B V_0 \int{d^d \! r  \lt(  U_{GL}  +  {|\vec{\nabla} \psi |^2 / 2} \rt)}
\eeq
where $U_{GL}=\eta^2 [ (T/T_0-1)  {| \tilde \psi({\bf r}) |}^2+ {{| \tilde \psi({\bf r}) |}^4/ 2} ]$  comes from the expansion of $S_0$ in powers $\beta \D$.
$T_0$ is the mean field pairing temperature, and   $ \tilde \psi = {\psi / \lt| \psi_0 \rt| }$ where $\psi_0 =\psi(T=0)$. $S_{GL}$ is normalised with a lattice spacing   $\varepsilon$. $ \eta:= \varepsilon / \xi_0$, where $\xi_0$ is the mean field correlation length at zero temperature. $V_0$ is the zero temperature phase stiffness.
The energy becomes
\beq
	E \approx S_0(\langle \D \rangle_{GL}) + \langle S_1 \rangle_{GL}
	\label{energy}
\eeq
where
$
S_0(\langle \D \rangle) = ({\langle \D \rangle^2 / U})  - {2 \over \beta V}\sum_q  \log
 [ 2 \cosh { \beta E_q/2} ]
$
corresponds the BCS free energy for which the gap value is determined by
the GL average. The $d$-wave quasi-particle energy is $ E_q = [ (\varepsilon_{q}-\mu)^2 +
 \langle \D \rangle^2 \cos^2(2\theta) ]^{1/2} $ where $\mu$ is the chemical potential.
 and $\theta$ is the angle in $k$ space with respect to $k_x$ direction.

For computer simulations of the statistical ensemble $\{\psi\}$ under the action
 $S_{GL}$, we use a standard Monte Carlo procedure to update amplitude $\R$ and a Wolff
\cite{wolff} algorithm for the phase $\phi$ as for the real $\Phi^4$ model
\cite{brower}. Typically $10^4$ sweeps are needed to obtain good statistics for lattice size: $N = 40^2$. The value of
$\eta$ depends on the corse-graining procedure and is fixed to 3 in this method.
Changes in $\eta$ modify slightly the shape of the average amplitude.
 The link between $S_0$ and the GL action is made by fixing ratio of $\Do$ and $T_0$. Since we are interested in
 the temperature domain from $T_c$ to $T^*$, the ratio $\Do/T_0$ is fixed at $T=T_c$ by requiring that the specific heat jump
  recovers its BCS value in the overdoped regime. We found that this ratio must be $\Do/T_0= 3.06$ in the $s$-wave case, and $\Do/T_0=3.72$ for $d$-wave symmetry.

Both approaches are valid below and above  the critical temperature $T_c$ which is the
temperature where the phase stiffness becomes zero.

\noindent
The {\bf specific heat} $C$ is the sum of  $C_0$ and  $C_1$, resp. amplitude and gradient contributions. Defining $\gamma = C/(\gamma_n T)$ where $\gamma_n$ the Sommerfeld constant,
the reduced specific heat is
\beq
\gamma = \gamma_0\lt(\langle \D \rangle \rt) + \gamma_1
\eeq
where $\gamma_1$ is divided by $T_c$ instead of $T$ since $S_1$ is classical
 and does not satisfy the third law of thermodynamics.
\begin{figure}[h]
\begin{center}
 \setlength\unitlength{1cm}
 \begin{picture}(8.5,5) \put(0,-0.5){\resizebox{8.5cm}{!}{\includegraphics{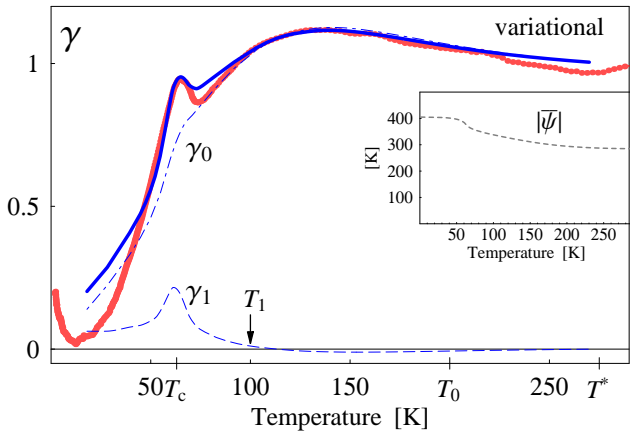}}}
 \end{picture}
\end{center}
\caption{ The specific heat from the variational method (thick),
 which is the sum of gradient (dashed) and amplitude (dotted-dashed) contributions,
 reproduces measurements  of YBa$_2$Cu$_3$O$_{6.73}$ (points). {\em Inset:} $T$ dependent amplitude $\Db$ from equation (\protect{\ref{amplitudeequation}}). }
\label{specificheat1}
\end{figure}
%
\begin{figure}[h]
\begin{center}
 \setlength\unitlength{1cm}
 \begin{picture}(8.5,5)
 \put(0,-0.5){\resizebox{8.5cm}{!}{\includegraphics{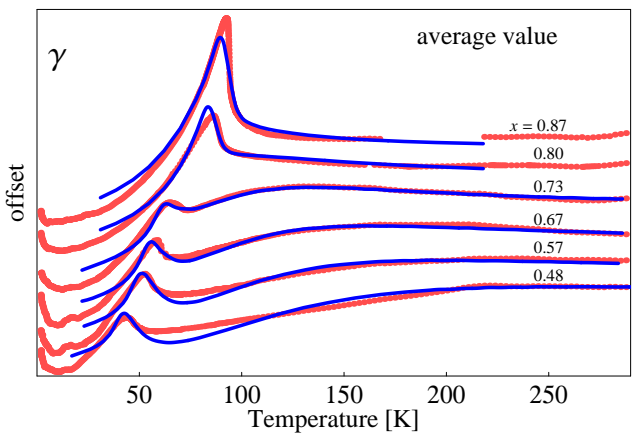}}}
 \end{picture}
\end{center}
\caption{Experimental specific heat (points) for YBa$_2$Cu$_3$O$_{6+x}$ compared to the average value method.}
\label{specificheat2}
\end{figure}
%
The amplitude contribution $\gamma_0$ is 1 at high temperature.
The gradient contribution is normalised as:
\beq
{C_1 \over \gamma_n} = {k_B \over \xi_0^3 \gamma_n} {C^{(s)}_1 \over N k_B}
\eeq
where $V=N \xi_0^3$, and $C^{(s)}_1/(N k_B)$ is the specific heat per number of lattice sites  coming from the simulations.  Experiments give $\gamma_n \approx$  26 mJ K$^{-1}$
mol$^{-1}$ = 252 J K$^{-1}$ m$^{-3}$. For the fit of Fig. \ref{specificheat1}, using the reasonable value $ \xi_0 \approx 16$ \AA,  we get  the dimensionless constant $\alpha={k_B /( \xi_0^3 \gamma_n}) \approx 13.5$.
In Fig. \ref{specificheat1} the experimental specific heat of YBa$_2$Cu$_3$O$_{6.73}$ \cite{loram94} is fitted using the variational method reproducing the double peak structure: a sharp peak at $T_c$ coming mainly from correlated phase fluctuations and a wide hump rounded by amplitude fluctuations (see Fig. \ref{fig-distribution} for typical distributions). The crossover temperature $T_1$  corresponds to the temperature where $\gamma_1$ is less than approximately 2\% of the  normal specific heat.
  In the amplitude equation (\ref{amplitudeequation}), a 2-dimensional density of states
$D(\varepsilon) = 1/W$ is used with $W=5000 \mbox{K}$, $\mu = 0.25 W$ and  $U=959 \mbox{K}$. 
These values give $T_0 \approx 200 \mbox{K}$ and $\psi_0 \approx 2.14 T_0$ in agreement with experiments \cite{kugler}.
 The other parameters are $V_0 = 72 \mbox{K}$ and $\eta=5$.

The average value method is used to reproduce specific heat measured for different dopings in Fig. \ref{specificheat2}. 
For underdoped systems $x<0.80$ we use simulations in $D=2$. For the more overdoped, $x\geq 0.80$, simulations are done in $D=3$.
  Parameters $V_0$  and $T_0$ extracted from the best fits are shown in the phase diagram of Fig. \ref{phasediagram2}. 
  $\alpha$ ranges from 7 to 15.
\begin{figure}[h]
\begin{center}
 \setlength\unitlength{1cm}
 \begin{picture}(8,5)
\put(0,-0.5){\resizebox{8cm}{!}{\includegraphics{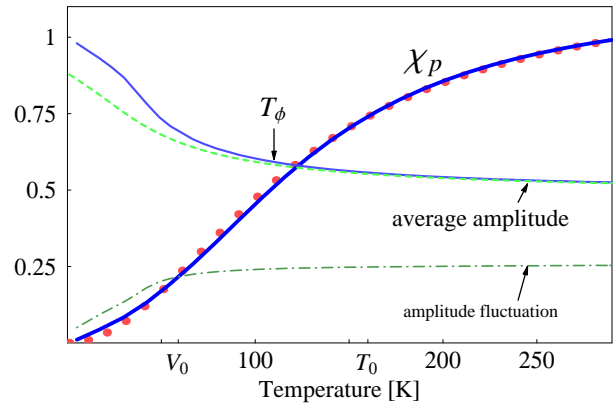}}}
\end{picture}
\end{center}
\caption{ \label{spinsusceptibility} The measured spin susceptibility  of
YBa$_2$Cu$_3$O$_{6.63}$ (points) divided by $\chi_0 $ is well fitted by the theoretical (thick) susceptibility $\chi_p$. 
 The dotted-dashed line is standard deviation of the average amplitude.}
\end{figure}
{\bf The magnetic susceptibility} $\chi$ has two contributions: the paramagnetic spin
susceptibility $\chi_p$ and the orbital diamagnetic susceptibility $\chi_d$.
$\chi_p$ has been measured by Takigawa {\em et al} \cite{takigawa}
on powder YBa$_2$Cu$_3$O$_{6.63}$  using Cu and O NMR experiments. The direct contribution coming from phases is negligible because NMR probes essentially the presence of pairs which is related to the amplitude. Therefore, $\chi_p$ is given by the amplitude contribution:
\beq
  {\chi_p} =  {\chi_0 \over 2 T}\int_0^{2 \pi} \! \! {d\theta \over 2 \pi}
\int_{0}^{\infty} \! \!
\! d\varepsilon  \
\cosh^{-2}{\sqrt{\varepsilon^2+ \langle \D \rangle_{GL}^2 \cos^2(2 \theta)} \over 2 T}
\label{chid}
\eeq
$\chi_0$ is the Pauli spin susceptibility. In Fig. \ref{spinsusceptibility}, we compare the result of equation (\ref{chid}) and Takigawa's measurements on powder YBa$_2$Cu$_3$O$_{6.63}$. The best fit yields $T_0 = 159.2 \mbox{K}$, $V_0 = 59.3\mbox{K}$. The competition
between amplitude and thermal energy enters into the spin susceptibility (\ref{chid}) by the ratio $\langle \D \rangle / T $. 
The temperature $T_\phi$ where phases start to influence the thermodynamics is the temperature where  $\langle \D \rangle$ deviates from the average amplitude computed for random phases.
 This temperature is found above $T_c$ at $T_\phi \approx 90\mbox{K}$.
 The orbital diamagnetic susceptibility $\chi_d$ in YBa$_2$Cu$_3$O$_{6.60}$ shows fluctuations effects up to about 15 K above $T_c$. For $T > T_{\phi}$, phase fluctuations are so strong that $\chi_d$ vanishes \cite{sewer}.\\
\noindent
 The {\bf Green function} can be calculated in CPA (Coherent Potential Approximation) for electrons that are scattered from spatially uncorrelated "pair impurities".
 We extended the $s$-wave CPA \cite{gyorffy} by using a $d$-wave amplitude  and uncorrelated phases. We have computed the density of states at half-filling $\mu = 0.5 W$ and maximum amplitude $\D = 0.8 W$.  The result is a pseudogap of width $\Delta \approx 0.2 W$ although phases are completely uncorrelated.\\
{\bf Phase diagram:} Values of $T_0$, $V_0$ and $T_\phi$ extracted from specific heat fits (Fig. \ref{specificheat2}) are reported in Fig. {\ref{phasediagram2}}.
 $T_1$ is the temperature where $C_1 / \gamma_n = 2\%$.
$T_\phi$ is computed in the same way as shown in Fig. \ref{spinsusceptibility} using average amplitudes from specific heat fits. Here, it is the temperature where the amplitude differs by $2\%$ from the amplitude in a random phase field.
The energy scale $E_g:=\langle |\psi|\rangle_{T=200\mbox{\footnotesize K}}$ of the pseudogap is defined as the amplitude at $T=200\mbox{K}$. $E_g$ shows the same doping dependence as the one found by Loram and Tallon \cite{loram2}. However it is not related to any hidden critical point.
 Phase correlations above $T_c$ grow rapidely in the underdoped regime following the $T_\phi$ line, an have a similar doping dependence as Nernst effect results \cite{wang}. The gradient specific heat from $S_1$ disappears more rapidely like in the  Hall effect \cite{matthey}.\\
%
%
\noindent
{\bf Discussion:} amplitude and phase fluctuations are the key for understanding the pseudogap regime of underdoped high temperature superconductors. Phase
correlations disappear completely near a temperature $T_{\phi}$ above $T_c$, and therefore,
for $T > T_{\phi}$, the pseudogap region is dominated by amplitude fluctuations.
The mean field temperature $T_0$ has a similar doping dependence as $T^*$, signaling that the pseudogap region is due to independent fluctuating pairs. Comparisons with measured specific heat on underdoped YBCO reproduce the double peak structure: a sharp peak at $T_c$ coming mainly from phase correlations and a separate wide hump below $T^*$ rounded by fluctuations. The spin susceptibility, related to the
amplitude, recovers its normal behaviour near $T^*$ whereas the orbital magnetic
susceptibility, related to phases, disappears near $T_{\phi}$. These considerations
are independent from the underlying pairing mechanism, and any microscopic theory inducing pairing 
should lead to similar conclusions.\\
 All these findings provide additional evidence for the fact that superconductivity and
pseudogap have the same origin. The former is primarily related to phases of the pairing
field, which are ordered below the transition temperature and whose correlations survive over a 
limited temperature region above $T_c$.
 The pseudogap regime of underdoped materials then extends to much higher temperatures due to the persisting amplitude fluctuations of the pairing field.\\
We thank J. Loram, B. Janko, C. Moca and C. Wirth. This work has been supported by the Swiss National Science Foundation.



\begin{thebibliography}{99}
\bibitem{loram94} J.W. Loram {\em et al}, Phys. Rev. Lett. {\bf 71}, 1740 (1993)


\bibitem{timusk} T. Timusk, S. Bryan, Rep. Prog. Phys. {\bf 62}, 61 (1999)

%

\bibitem{emery} V.J. Emery, S.A. Kivelson, Nature {\bf 374}, 434-437 (1995)

\bibitem{randeria} M. Randeria, J.C. Campuzano, cond-mat/9709107

\bibitem{schmalian} J. Schmalian {\em et al}, Phys. Rev. Lett. {\bf 80}, 3839, (1998)

\bibitem{lee} P. A. Lee, Physica C {\bf 317-318}, 194 (1999)

\bibitem{sachdev} S. Sachdev, Physics World, 12(4), 33 (1999)

\bibitem{takigawa} M. Takigawa {\em et al}, Phys. Rev. B {\bf 43}, 247 (1991)

\bibitem{demsar} J. Demsar {\em et al}, Phys. Rev. Lett. {\bf 82}, 4918 (1999)
\bibitem{matthey} D. Matthey {\em et al}, Phys. Rev. B {\bf 64}, 24513 (2001)




\bibitem{BCS} J. Bardeen {\em et al}, Phys. Rev. {\bf 108}, 1175 (1957)

\bibitem{curty} Ph. Curty, H. Beck,  Phys. Rev. Lett. {\bf 85}, 796 (2000)

\bibitem{wolff} U. Wolff,  Phys. Rev. Lett. {\bf 62}, 361 (1989)
\bibitem{brower} R.C. Brower {\em et al}, Phys. Rev. Lett. {\bf 62}, 1087 (1989)

\bibitem{kugler} M. Kugler {\em et al}, Phys. Rev. Lett. {\bf 86}, 4911 (2001)
\bibitem{sewer} A. Sewer, H. Beck, Phys. Rev. B {\bf 64}, 14510 (2001)

\bibitem{gyorffy} B.L. Gyorffy {\em et al}, Phys. Rev. B {\bf 44}, 5190 (1991)


\bibitem{wang} Y. Wang {\em et al}, Phys. Rev. B {\bf 64}, 224519 (2001)

\bibitem{loram2} J.L. Tallon, J.W. Loram, cond-mat/0005063

\end{thebibliography}
\end{document}